\documentclass[aps,pra,reprint,superscriptaddress]{revtex4-2}

\usepackage{graphicx}
\usepackage{dcolumn}
\usepackage{bm}
\usepackage{amsmath,amssymb}
\usepackage{amsthm}
\usepackage{tikz}
\usepackage{braket}
\usepackage{quantikz}
\usepackage{xcolor}
\colorlet{darkgreen}{green!35!black}
\usepackage{url}
\usepackage{float}
\usepackage{booktabs}

\usepackage{listings}
\usepackage{xcolor}
\usepackage{caption}
\definecolor{codegray}{gray}{0.95}
\lstdefinestyle{mystyle}{
    backgroundcolor=\color{codegray},   
    commentstyle=\color{gray},
    keywordstyle=\color{blue},
    numberstyle=\tiny\color{gray},
    stringstyle=\color{orange},
    basicstyle=\ttfamily\small,
    breaklines=true,                 
    captionpos=b,                    
    keepspaces=true,                 
    numbers=left,                    
    numbersep=5pt,                  
    showspaces=false,                
    showstringspaces=false,
    showtabs=false,                  
    tabsize=4
}
\usepackage{hyperref}
\usepackage{color}

\begin{document}

\title{Context-Dependent Time--Energy Uncertainty Relations from Projective Quantum Measurements}


\author{Mathieu Beau}
\affiliation{Department of Physics, University of Massachusetts, Boston, Massachusetts 02125, USA}

\date{\today}

\begin{abstract}
We introduce a general framework for defining context-dependent time distributions in quantum systems using projective measurements. The time-of-flow (TF) distribution, derived from population transfer rates into a measurement subspace, yields a time--energy uncertainty relation of the form \( \Delta \mathcal{T} \cdot \Delta H \geq \hbar / (6\sqrt{3}) \times \delta\theta \), where \( \delta\theta \) quantifies net population transfer measured by the projector. This bound applies to arbitrary projectors under unitary dynamics and reveals that time uncertainty is inherently measurement-dependent. We demonstrate the framework with two applications: a general time-of-arrival (TOA)--energy uncertainty relation and a driven three-level system under detuned coherent driving. The TF framework unifies timing observables across spin, atomic, and matter-wave systems, and offers an experimentally accessible route to probing quantum timing in controlled measurements.
\end{abstract}

\maketitle

\section{Introduction.} Time in quantum mechanics defies a universal definition. Unlike position or energy, it lacks a canonical operator and is not treated as an observable in the standard formalism~\cite{pauli1933handbuch}. Instead, time typically enters as an external parameter, which obscures its role in processes such as state transitions, arrival events, or dynamical flows~\cite{Muga1,Muga2}. Conceptual approaches such as the Page--Wootters mechanism~\cite{Wootters83} and quantum clock models~\cite{Giovannetti15,Maccone20}, which infer time from correlations with auxiliary systems, offer elegant solutions in principle, but remain largely unexplored experimentally.

Time also plays a central role in modern quantum technologies, where precise timing is essential for quantum control, decoherence, and finite-time thermodynamic processes~\cite{WisemanMilburn2010,Chenu17,Beau17,Chen10,An16,delCampo19,Watanabe17}. In this context, quantum speed limits (QSLs)~\cite{mandelstam1945uncertainty,margolus1998maximum} provide lower bounds on the time required for a transition, and have found applications in optimal control and open-system analysis~\cite{Deffner17,delCampo13}. However, QSLs offer only geometric or energetic constraints, they do not describe the statistical distribution of transition times, nor are they directly linked to measurement outcomes.

{Several approaches have been proposed to define quantum time distributions, in particular for the long-standing time-of-arrival (TOA) problem (see reviews \cite{Muga1,Muga2}). Semiclassical or hybrid methods randomize classical trajectories under initial uncertainty \cite{udovic1993neutron,kurtsiefer1995time,kothe2013time,GBAR14,gliserin2016high,GBAR22}; operator-based approaches introduce non-self-adjoint or POVM-defined TOA operators \cite{Kijowski74,Werner86}; self-adjoint TOA operators have been constructed in specific contexts \cite{Rovelli96,Galapon02}; Bohmian formulations identify the TOA with the absolute value of the probability current \cite{leavens1993arrival,Mckinnon1995,leavens1998time}; and detector--clock models reveal dynamical limits to arrival-time measurements \cite{Aharonov98}. However, semiclassical/hybrid schemes neglect fully quantum interference and backflow; operator/POVM proposals often lack explicit, broadly applicable experimental protocols and are typically worked out for free or asymptotically free dynamics; self-adjoint constructions hold only for special Hamiltonians (e.g., discrete/semibounded spectra) and do not extend generally; flux/current definitions face sign and normalization issues (e.g., backflow) and are tied to continuous position observables; and detector--clock models suffer from measurement back-action and Zeno/reflective effects. In discrete systems, quantum trajectories~\cite{Gambetta08} provide timing information by unraveling open-system dynamics into stochastic jumps or diffusive paths, but they rely on continuous monitoring and are affected by the quantum Zeno effect. First-detection protocols~\cite{Liu20} extract statistics from stroboscopic projective measurements, yet they emphasize the first detection event and remain invasive. What remains missing across these approaches is a unifying and operational framework that recovers the flux-based TOA distribution in continuous systems, extends naturally to discrete spectra, and applies consistently to both closed and open dynamics for arbitrary Hamiltonians.
}

In this work, we propose a unified and operational framework for quantum timing observables based on the concept of a \emph{Time-of-Flow} (TF) distribution, which extends naturally to both discrete systems and matter-wave dynamics. 
The TF distribution quantifies the temporal profile of population change into a chosen subspace of the Hilbert space, be it a discrete energy level or a spatial detection region. It captures both transition timing in finite-level systems and the time-of-arrival (TOA) for particles moving in space. The key advantage of the TF framework is that it avoids the need for idealized continuous measurements or auxiliary clock systems. Instead, it is constructed from projective measurements performed on an ensemble at discrete times, forming a meta-protocol that is both measurement-agnostic and experimentally implementable, and that bypasses the quantum Zeno effect for continuous measurement protocols \cite{Wineland90}. Once reconstructed, the TF distribution defines a genuine time random variable \( \mathcal{T} \), allowing for the computation of statistical moments such as the mean and variance, even when time is not represented as an operator. Unlike conventional observables such as position or momentum, time in our framework is defined contextually, as a statistical construct derived from projective measurement outcomes. This approach avoids the need for a time operator and instead grounds timing in experimentally accessible procedures.

This approach leads to a general and operational time--energy uncertainty relation of the form
\begin{equation}
    \Delta \mathcal{T} \cdot \Delta H \geq \frac{\hbar}{6\sqrt{3}} \times \delta\theta,
\end{equation}
where \( \Delta \mathcal{T} \) is the standard deviation of the TF distribution, \( \Delta H \) is the energy spread, and \( \delta\theta \) quantifies the total population transfer into the measurement subspace. Unlike geometric bounds from quantum speed limits (QSLs), this inequality captures the full temporal spread of a measurable probability distribution and depends explicitly on the measurement protocol. It thus reflects the context-dependent nature of time uncertainty in quantum mechanics, reinforcing the idea that time, as inferred from data, is not a universal observable but a relational quantity tied to a specific measurement channel.

The generality of our method allows for its application to a broad range of systems, from driven three-level atoms to free-falling cold atomic clouds. In particular, we show how the TF distribution recovers the standard TOA distribution as a special case, with the time-of-arrival uncertainty constrained by the energy variance of the particle. The framework is experimentally accessible and can be implemented in platforms such as ultracold atom interferometry, optical lattices, and superconducting circuits, providing a versatile tool to probe quantum timing beyond conventional paradigms, directly from a measurable statistical distribution.

\begin{figure*}[ht]
    \centering
    \includegraphics[width=0.8\linewidth]{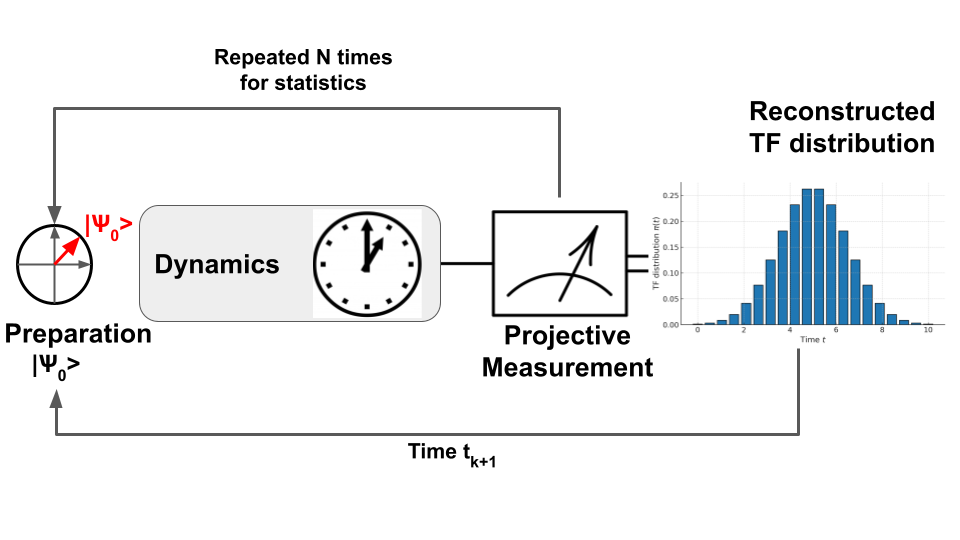}
    \caption{\textbf{Schematic of the Protocol for Reconstructing the TF Distribution.} After preparing the initial state and letting the system evolve, a projective measurement $\hat{M}$ is performed at a chosen time $t_k$; detection/non-detection is recorded and the run is repeated to build statistics. This loop is then repeated for a grid of times $\{t_k\}_{k=1}^n$ to reconstruct the detection–time profile $p(t_k)$ and, from its finite-time differences, the TF distribution. This single-shot, time-gated scheme avoids Zeno effects and provides a unified framework for observables with continuous or discrete spectra.}
    \label{Fig:Protocol}
\end{figure*}


\section{Time-of-Flow (TF) distribution for a projective measurement.}

In quantum theory, a projective measurement is described by a projection onto a subspace \( \mathcal{M} \) of the Hilbert space. The corresponding projector \( \hat{M} \) satisfies \( \hat{M}^2 = \hat{M} \) (idempotency) and \( \hat{M}^\dagger = \hat{M} \) (Hermiticity), ensuring it represents a valid observable. The probability of detecting the system within \( \mathcal{M} \) at time \( t \) is given by  
\(
p(t) = \mathrm{Tr}[\hat{\rho}_t \hat{M}],
\)  
where \( \hat{\rho}_t \) is the density matrix of the system at time \( t \).

We are interested in the temporal variation of this probability to gain insight into the rate of population change within the measurement subspace. This rate is defined as  
\(
r(t) = \frac{d}{dt}p(t) = \mathrm{Tr}\left[ \mathcal{L}(\hat{\rho}_t)\hat{M} \right],
\)  
where \( \mathcal{L}(\cdot) \) denotes the Liouvillian superoperator governing the system’s dynamics \cite{Breuer2002,lidar2019lecture}. The sign of \( r(t) \) indicates whether population is flowing into (\( r(t)>0 \)) or out of (\( r(t)<0 \)) the subspace \( \mathcal{M} \), similar to a current across a boundary in space.

To capture the magnitude of this population flow, irrespective of its direction, we define the instantaneous flow rate as \( d(t) = |r(t)| \). Normalizing this quantity over a finite time interval, we introduce the \textit{Time-of-Flow (TF) distribution}:
\begin{equation}\label{Eq:TFdistrib:General}
\pi(t) = \mathcal{N} \times \left| \frac{d}{dt}p(t) \right| = \mathcal{N} \times \left| \mathrm{Tr}\left[ \mathcal{L}(\hat{\rho}_t) \hat{M} \right] \right|,
\end{equation}
{with normalization constant
\(
\mathcal{N}^{-1} = \int_0^{t_f} \left| \frac{d}{dt}p(t) \right| dt
\), where $t_f$ is the final time targeted by the experiment. }

{Note that the final time $t_f$ is chosen according to the purpose of the analysis. For instance, if the goal is to maximize the quantum transition between two states, one selects $t_f$ at the time of maximal occupation of the target state, as discussed in Section~\ref{Section:3level}. In contrast, for the time-of-arrival problem in matter-wave dynamics, $t_f$ is taken to be as large as possible (ideally $+\infty$) to ensure that the particle is detected; see Section~\ref{Section:TOA}.}
The TF distribution thus characterizes the probability density for a change of occupation to occur within the measurement subspace at time \( t \). In stationary regimes, \( r(t) = 0 \) and \( \pi(t) = 0 \); when the system evolves, \( \pi(t) \) quantifies the temporal profile of the transition process.
This concept was introduced in Ref.~\cite{Beau25_2} for discrete quantum systems. Here, we generalize the formalism to encompass both discrete and matter-wave dynamics, making it broadly applicable to unitary and open dynamics.
  
The question now is: how can we measure the TF distribution \( \pi(t) \), or more precisely, what is the experimental protocol to reconstruct it? We propose the following procedure, a general \textit{meta-protocol} that applies to any projective measurement, designed to avoid Zeno inhibition by relying on ensemble measurements at discrete times (See Figure \ref{Fig:Protocol}):
\begin{enumerate}
    \item \textbf{State preparation:} Prepare the system in a known initial state \( \hat{\rho}_0 \) and let it evolve under its natural dynamics (unitary or open).

    \item \textbf{Measurement:} At a chosen time \( t_k \), perform a projective measurement defined by the operator \( \hat{M} \). Record whether the system is detected in the corresponding subspace.

    \item \textbf{Statistics:} Repeat the procedure (steps 1-2) many times for \( t\ =\ t_k \) to obtain the detection count \( N_k \).

    \item \textbf{Temporal scan:} Repeat steps 1–3 for a sequence of times \( \{t_1, t_2, \ldots, t_n\} \) with time step \( \delta t \) small enough to resolve temporal features. The normalized detection probabilities 
    \[ p(t_k) = \dfrac{N_k}{\sum_{l=1}^n N_l} \] 
    approximate \( \mathrm{Tr}[\hat{\rho}_{t_k} \hat{M}] \).

    \item \textbf{TF distribution:} The discrete approximation to the TF distribution is given by
    \[
    \pi(t_k) = \frac{1}{\delta t} \times \frac{|N_{k+1} - N_k|}{\sum_{j=1}^{n-1} |N_{j+1} - N_j|},\ k=1,\cdots,n-1.
    \]
    A histogram of \( \pi(t_k) \) provides an experimental reconstruction of the TF distribution.
\end{enumerate}

This protocol extends the one introduced in Refs.~\cite{Beau24,Beau24_2} for time-of-arrival (TOA) distributions. It generalizes the approach to arbitrary projective measurements and, as we will show later, reproduces the TOA distribution as a special case when the projector corresponds to spatial detection. For a more detailed analysis of the TF approach in discrete systems, particularly the optimization of two-level protocols and their connection to quantum speed limits (QSLs), see the companion paper \cite{Beau25_TFEntropy}.  \\

\section{Lower bound for the time-uncertainty.} Let \( \mathcal{T} \) be a random variable associated with the TF distribution. Note that in this framework, time is not a direct observable but a derived quantity, constructed from the statistics of projective measurements. In this sense, time is not a universal observable like position, but a context-dependent quantity determined by the measurement protocol. We can easily calculate the moments of \( \mathcal{T} \) using the standard formula \( \langle \mathcal{T}^n \rangle = \int_0^\infty dt\, t^n \pi(t) \). The spread of the distribution is characterized by its standard deviation \( \Delta \mathcal{T} = \sqrt{\langle \mathcal{T}^2 \rangle - \langle \mathcal{T} \rangle^2} \).  
By Chebyshev's inequality, for the random variable \( \mathcal{T} \) with distribution \( \pi(t) \) and finite variance \( (\Delta \mathcal{T})^2 \), we have
\[
\text{Prob}\left( |\mathcal{T} - \langle \mathcal{T} \rangle| \geq k\Delta \mathcal{T} \right) = \int_{|t - \langle \mathcal{T} \rangle| \geq k\Delta \mathcal{T}} \pi(t)\, dt \leq \frac{1}{k^2}\ ,
\]
where $\ k\geq 1$, \( \mathcal{T} \) is the random variable associated with the TF distribution, \( \langle \mathcal{T} \rangle \) its mean value, and \( \Delta \mathcal{T} \) its standard deviation. 
Thus,  the probability mass located inside the interval
\(
\Omega_k \equiv [\langle \mathcal{T} \rangle - k\Delta \mathcal{T},\, \langle \mathcal{T} \rangle + k\Delta \mathcal{T}]
\)
must be at least \(  (1 - k^{-2}) \). 
Since the maximum value of the distribution is \( \pi_{\text{max}} = \max_{t \in \Omega_k} \pi(t) \), the probability contained in any interval of width \( 2k\Delta \mathcal{T} \) is bounded above by
\(
\pi_{\text{max}} \times 2k\Delta \mathcal{T}.
\)
Therefore, the probability inside the \( 2k\Delta \mathcal{T} \) interval centered at \( \langle \mathcal{T} \rangle \) satisfies {
\(
\text{Prob}(\mathcal{T} \in [\langle \mathcal{T} \rangle - k\Delta \mathcal{T},\, \langle \mathcal{T} \rangle + k\Delta \mathcal{T}]) \leq 2k\Delta \mathcal{T}\times \pi_{\text{max}}.
\)}
But we already know this probability must be at least \( 1 - k^{-2} \) (from Chebyshev's inequality), so we obtain the lower bound:
\begin{equation}\label{Eq:lowerbound:DeltaTF}
\Delta \mathcal{T} \geq \frac{1}{3\sqrt{3}} \times \frac{1}{\pi_{\text{max}}} \geq \gamma \times \frac{1}{\pi_{\text{max}}}, 
\end{equation}
where \( \gamma = (k^2 - 1)/(2k^3),\ k \geq 1 \), and the sharpest bound is obtained for \( k = \sqrt{3} \), which yields \( \max_{k\geq 1}(\gamma) = (3\sqrt{3})^{-1} \).
This inequality shows that the standard deviation \( \Delta \mathcal{T} \) of the TF distribution cannot be made arbitrarily small if \( \pi_{\text{max}} \) is bounded, enforcing a minimal temporal spread inversely proportional to the peak height of the TF distribution. 

It is also worth noting that this bound is \textit{context-dependent}, in that \( \pi_{\text{max}} \), and thus the minimal temporal spread, depends explicitly on the choice of the measurement operator \( \hat{M} \) and the system’s dynamics \( \hat{\rho}_t \). This reflects the fact that time uncertainty is not a universal property, but rather emerges from the specific measurement context.\\

\begin{figure*}[ht!]
    \centering
    \includegraphics[width=1.0\linewidth]{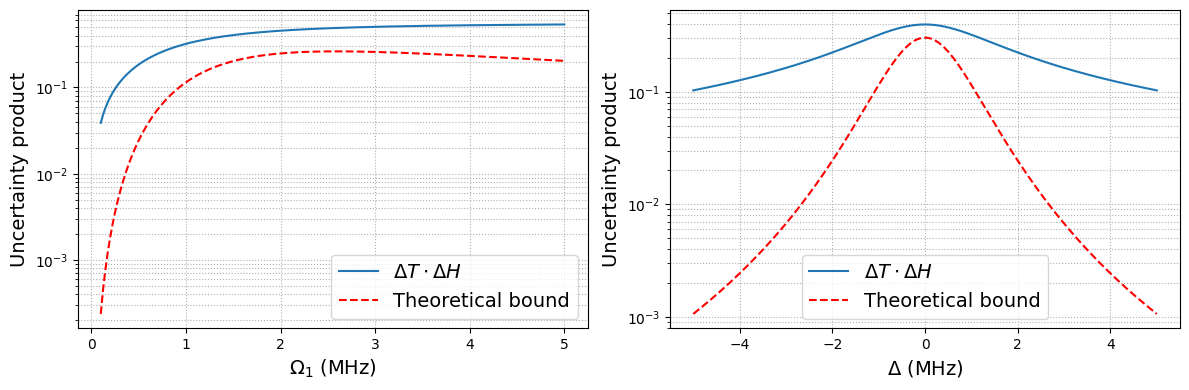}
    \caption{\textbf{Time–energy uncertainty product \(\Delta \mathcal{T} \cdot \Delta H\) and its theoretical lower bound for a driven three-level system.} 
On the left panel, we show the uncertainty product (solid blue) and bound (dashed red) as functions of the Rabi frequency \(\Omega_1\), with fixed \(\Omega_2 = 1\) MHz and detuning \(\Delta = 1\) MHz. 
On the right panel, we plotted the same quantities versus detuning \(\Delta\), with fixed \(\Omega_1 = \Omega_2 = 1\) MHz. 
In both panels, the uncertainty product respects the bound (see dashed line) \(\Delta \mathcal{T} \cdot \Delta H \geq \frac{\hbar}{6\sqrt{3}} \cdot \delta\theta\), with \(\delta\theta = |p(t_f) - p(0)|\) the net transition probability. The standard deviation $\Delta \mathcal{T}$ was calculated in the set time range to one Rabi oscillation period $T=2\pi/\Omega$, where $\Omega=\sqrt{\Omega_1^2+\Omega_2^2+\Delta^2}$. }
    \label{Fig:Rabi}
\end{figure*}

\section{Time–energy uncertainty relation for closed quantum systems.}\label{Section:TEUR}
We consider a closed quantum system for which the dynamics is governed by the Heisenberg equation
\begin{equation}\label{Eq:Master:CLosedSystem}
    \frac{d}{dt}\hat{\rho}_t = \mathcal{L}(\hat{\rho}_t) = -\frac{i}{\hbar}[\hat{H},\hat{\rho}_t]\ ,
\end{equation}
where \( \hat{\rho}_t \) is the density matrix and \( \hat{H} \) is a time-independent Hamiltonian.

From equation \eqref{Eq:TFdistrib:General}, we obtain the uniform bound in time:
\begin{equation}\label{Eq:UniformBoundTFdistrib}
    \pi(t)\leq \frac{2\Delta H}{\hbar\, \delta\theta}\ , 
\end{equation}
where \( \delta\theta \equiv |p(t_f)-p(0)| \) quantifies the net population transfer, and \( \Delta H \equiv \sqrt{\langle \hat{H}^2 \rangle - \langle \hat{H} \rangle^2} \) is the standard deviation of the energy {with respect to the initial state, where $\langle \hat{H}^n \rangle \equiv \mathrm{Tr}\left(\hat{H}^n\hat{\rho}_0\right),\ n\in\mathbb{N}$}. This inequality follows from the estimate \( |\mathrm{Tr}([\hat{H},\hat{\rho}_t] M)| \leq 2\Delta H \) and the normalization bound \( \mathcal{N}^{-1} \geq \delta\theta \) (see \cite{SM} for details). Combining this with the general lower bound from equation \eqref{Eq:lowerbound:DeltaTF}, we obtain
\begin{equation}\label{Eq:TimeUncertaintyRelation:General}
    \Delta \mathcal{T} \cdot \Delta H \geq \frac{\hbar}{6\sqrt{3}} \times \delta\theta\ .
\end{equation}

To our knowledge, this is the first time–energy uncertainty relation that involves a genuine statistical measure of time, i.e., the standard deviation \( \Delta \mathcal{T} \), rather than a heuristic or geometric time scale as appears in conventional quantum speed limits (QSLs).

This inequality can be interpreted as a Heisenberg-like uncertainty relation between the energy observable \( \hat{H} \) and the random variable \( \mathcal{T} \), which represents the timing of a specific projective measurement \( \hat{M} \). Notably, \( \mathcal{T} \) refers to an external, operational time determined by the measurement protocol, rather than to an intrinsic or internal time parameter.

{The TF standard deviation $\Delta \mathcal{T}$ in equation \eqref{Eq:TimeUncertaintyRelation:General} should be interpreted as the temporal spread of the transition or arrival process under a given measurement context. In practice, the Time-energy uncertainty relation (TEUR) inequality shows that $\Delta \mathcal{T}$ plays the role of a minimal uncertainty window: for fixed energy dispersion $\Delta H$, the timing of the event cannot be confined more sharply than this spread. This complements standard quantum speed limits, which constrains on transfer times, by instead constraining the distributional width. Thus $\Delta \mathcal{T}$ provides an operational lower bound on how precisely one can schedule or resolve quantum transitions, whether for state transfer in a discrete system or for time-of-arrival in continuous motion. In addition, the TF distribution itself provides access to the \emph{average transition time} (for discrete systems) or the \emph{average time of arrival} (for continuous systems), while $\Delta \mathcal{T}$ quantifies the spread of realizations around this average. From an engineering perspective, this means that one can tailor protocols to sharpen the distribution if a well-defined, sudden transition is desired, or allow for a broader distribution if a more gradual or robust process is advantageous.
}

For the more general case of a time-dependent Hamiltonian \( H = H(t) \), the bound \eqref{Eq:UniformBoundTFdistrib} can be replaced by its supremum over time, \( \max_t \Delta H(t) \). In this case, the uncertainty relation \eqref{Eq:TimeUncertaintyRelation:General} holds with \( \Delta H \) replaced by \( \max_t \Delta H(t) \). {In the companion paper \cite{Beau25_TFEntropy}, we also derive an alternative time--energy uncertainty relation in which the standard deviation of the Hamiltonian,
$
\Delta_k H \equiv \sqrt{\mathrm{Tr}\!\left(\hat{H}^2\hat{M}_k\right)-\mathrm{Tr}\!\left(\hat{H}\hat{M}_k\right)^2},
$
is computed with respect to the projector onto a single state \( \hat{M}_k \equiv \ket{k}\bra{k} \) rather than the time-dependent density matrix \( \hat{\rho}_t \). This formulation extends naturally to open quantum systems and has practical relevance for estimating the spread of the distribution for transitions between two quantum states in the discrete case. However, it does not capture the relationship between the energy dispersion
$
\Delta H \equiv \sqrt{\mathrm{Tr}\!\left(\hat{H}^2\hat{\rho}_0\right)-\mathrm{Tr}\!\left(\hat{H}\hat{\rho}_0\right)^2}
$
of the system and the temporal spread. Moreover, the standard deviation of the Hamiltonian with respect to a projector \( \hat{M} \) may fail to converge when the energy spectrum is continuous (e.g., for matter-wave dynamics, as in Sec.~\ref{Section:TOA}). For this reason, the present bound \eqref{Eq:UniformBoundTFdistrib} provides a unifying framework that applies to both discrete and continuous time control.}

\section{Time of arrival for closed quantum systems.}\label{Section:TOA} 
Consider a quantum system where $\mathcal{L}(\hat{\rho}_t) = -(i/\hbar)[\hat{H},\hat{\rho}_t]$ with $\hat{H}=\hat{p}^2/(2m)+V(\hat{x})$, where $\hat{p}$ is the momentum operator, $V(x)$ is a position-dependent potential, and the half–space projector $\hat{M}_x=\int_{-\infty}^x dx'\ket{x'}\bra{x'}$. Then, we find that the TF distribution is nothing but the time of arrival (TOA) distribution at a given position $x$ \cite{Beau24_2}
\begin{equation}\label{Eq:TOAdistr}
    \pi_x(t) = \dfrac{|j(x,t)|}{\int_0^{\infty} dt\ |j(x,t)|}\ ,
\end{equation}
as the quantum current of the particle $j(x,t)  = \text{Tr}(\hat{\rho}_{t} \hat{J}(x))$, where the surface-current operator $\hat{J}(x)=\frac{1}{2m}(\hat{p}\ket{x}\bra{x}+\ket{x}\bra{x}\hat{p})$, and as the denominator in \eqref{Eq:TFdistrib:General} $-(i/\hbar)\text{Tr}([\hat{H},\hat{\rho}_{t}] \hat{M})=-(i/\hbar)\text{Tr}(\hat{\rho}_{t} [\hat{M}_x,\hat{H}])=-\text{Tr}(\hat{\rho}_{t} \hat{J}(x))$ (we took the final targeted time of the experiment $t_f=+\infty$, assuming the integral in \eqref{Eq:TOAdistr} converges). To prove this formula, we used the identity $-(i/\hbar)[\hat{H},\hat{M}_x]=\hat{J}(x)$, see proof in \cite{SM}. The formula \eqref{Eq:TOAdistr} means that measuring the flow rate of the projective meaurement $\hat{M}_x$ is equivalent of measuring the mean value of the surface-current operator $\hat{J}(x)$. The TF distribution helps understanding why equation \eqref{Eq:TOAdistr} can be interpreted as the TOA distribution at a given position $x$. The TF distribution gives the normalized value of the rate of change of probability for the particle to enter (if $dp(t)/dt >0$) or to leave (if $dp(t)/dt <0$) the region $(-\infty,x]$. Hence, it measures the probability for the particle to pass through the position $x$, explaining why we obtain the absolute value of the current in equation \eqref{Eq:TOAdistr}. Thus, instead of measuring the cumulative distribution $F_t(x)=\mathrm{Tr}(\hat{\rho}_t \hat{M}_x)$ via the projector $\hat{M}_x$ as described by the meta-protocol above, formula \eqref{Eq:TOAdistr} provides an alternative protocol, consisting on a more direct measurement of the TOA distribution, based on the measurement of the quantum current (or its absolute value to be more precised) as previously proposed in \cite{Beau24,Beau24_2,Beau25_2} and common practice in experimental scenarios involving cold atoms \cite{Dalibard95,Dalibard96,Dalibard96_2} and quantum gases \cite{Boiron06, Yasuda96, Greiner05}. {We emphasize that while the current-based formula for TOA has long existed in the literature~\cite{Kijowski74,Werner86,Muga1,Muga2, leavens1998time}, our framework provides (i) an explicit experimental protocol that operationally grounds this expression, and (ii) a transparent interpretation of TOA as the rate of probability flow, and (iii) a direct proof that the current formula follows naturally from this construction.}

From the general time-energy uncertainty relation \eqref{Eq:TimeUncertaintyRelation:General}, we find the following TOA-energy uncertainty relation
\begin{equation}\label{Eq:TOA-energyUncertainty}
    \Delta T_x \Delta H \geq \frac{\hbar}{6\sqrt{3}} \times F_0(x)\ ,
\end{equation}
where $F_t(x) = \int_{-\infty}^x dx'\ \rho_t(x')$ is the probability to find the particle in the region $x'<x$ and where we found $\delta\theta = |F_{\infty}(x)-F_0(x)|=F_0(x)$, assuming $F_{\infty}(x)=0$ (if the targeted time $t_f<+\infty$, we should keep $\delta\theta$ in the numerator of \eqref{Eq:TOA-energyUncertainty}). This relation implies a trade-off: greater precision in the energy of the system results in increased uncertainty in the time of arrival (TOA). {Note that the analog bound in the time--energy relation derived in \cite{Beau25_TFEntropy}, for which the standard deviationof the Hamiltonian is defined as $\Delta_x H \equiv \sqrt{\text{Tr}\left(\hat{H}^2\hat{M}_x\right)-\text{Tr}\left(\hat{H}\hat{M}_x\right)^2}$, tends to zero as the trace of the commutator \( [\hat{H}, \hat{M}_x] \) diverges, thus the bound in equation \eqref{Eq:TOA-energyUncertainty} is more suitable for operators with continuous spectrum (e.g., position or momentum in matter-wave dynamics). }

As an interesting application, we consider the free fall of an atom in a constant gravitational field where $\Delta H = mg\sigma\sqrt{1+\frac{\hbar^4}{32g^2m^4\sigma^6}}$ \cite{Beau25_2}. If we consider the far-field regime $x\gg \sigma$, which is often the case experimentally, the cumulative distribution $F_0(x)\approx 1$, hence from the bound in equation \eqref{Eq:TOA-energyUncertainty}, we obtain 
\begin{equation}\label{Eq:DeltaTOA:lowerbound}
\Delta T_x \geq \dfrac{\hbar}{6\sqrt{3}\Delta H}\ .
\end{equation}
Note that for $\sigma \geq \sigma_c = \left(\dfrac{\hbar^2}{4\sqrt{2}m^2 g}\right)^{1/3}$, the standard deviation of the energy is dominated by the potential energy contibution $\Delta H \leq \sqrt{2}\ mg\sigma$, so we have a TOA uncertainty bounded below by 
\begin{equation}\label{Eq:DeltaTOA:lowerbound:SpecialRegime}
\Delta T_x \geq \dfrac{\sqrt{2}}{6\sqrt{3}}\delta\theta\times\frac{\hbar}{2mg\sigma}
\end{equation}
This means that $\Delta T_x$ decreases as $\sigma$ becomes large compared to $\sigma_c$, consistent with the particle exhibiting a more semiclassical behavior, causing the TOA distribution to become increasingly peaked around its classical arrival time. A similar but sharper bound $\Delta T_x\geq \hbar/(2mg)$ was found in \cite{Beau25_2}, up to the factor $(1/6)\sqrt{(2/3)}\ \delta\theta$, for some limiting cases and numerically for more general regimes, but the relation was not proven rigorously as we did here in the regime $\sigma\geq \sigma_c$. Further investigation will be presented in a separate paper to determine whether this relation holds universally or is valid only in specific regimes.

\begin{table}[h!]
\begin{tabular}{lcccc}
\toprule
\textbf{Atoms} & $\mathrm{\overline{H}}$ & K-39 & Rb-87 & Cs-133 \\
\midrule
$\sigma_c$ ($\mu$m) & 4.16 & 0.36 & 0.21 & 0.16 \\
$\Delta T_x \geq$ ($\mu$s) & 8.62 & 15.96 & 7.18 & 4.68 \\
\bottomrule
\end{tabular}
\caption{\textbf{Time-of-arrival uncertainty $\Delta T_x$ for various atomic species using the bound \eqref{Eq:DeltaTOA:lowerbound}}. The values represent the highest time resolution required to detect quantum effects in time-of-arrival measurements under free fall in Earth's gravity.}
\label{Table:toa_atoms}
\end{table}

In Table~\ref{Table:toa_atoms}, we show the minimum time-of-arrival uncertainty \( \Delta T_x \) for various atomic species, including antihydrogen atom (see in particular the GBAR experiment \cite{GBAR14,GBAR19,GBAR22,GBAR22_2}), Potassium-39 and Rubidium-87, and Cesium-133, derived from the position--TOA uncertainty bound~\eqref{Eq:DeltaTOA:lowerbound}. The critical spatial width \( \sigma_c \) is defined from the condition where the bound \eqref{Eq:DeltaTOA:lowerbound:SpecialRegime} is valid. The corresponding values of \( \Delta T_x \) represent the highest temporal resolution required to resolve quantum effects in TOA measurements under free fall in Earth's gravity. These estimates and our bounds may serve as practical benchmarks for experiments with cold atoms~\cite{Dalibard95,Dalibard96,Dalibard96_2}, free-falling matter waves~\cite{QuantumTest14}, and microgravity platforms~\cite{MicrogravityEarth10,MicrogravityEarth13}, including QUANTUS and MAIUS missions~\cite{MicrogravityEarth13,MAIUS18}, and the Bose-Einstein Condensate and Cold Atom Laboratory (BECCAL)~\cite{CAL21,CAL22,CAL23}. They are also relevant for future initiatives such as the Gravitational Behaviour of Anti-hydrogen at Rest (GBAR) experiment~\cite{GBAR14,GBAR19,GBAR22,GBAR22_2} and the Space-Time Explorer and Quantum Equivalence Principle Space Test (STE-QUEST)~\cite{Altschul15,ExpReview21}.

\section{Three-Level System with Detuned Rabi Coupling}\label{Section:3level}

To illustrate the application of the time-of-flow framework and our time-energy uncertainty relation to non-trivial discrete systems, we consider a three-level system with coherent driving and detuning. The Hamiltonian in the rotating wave approximation is given by \cite{Shore1990,Bergmann1998}
\begin{equation}\label{Eq:Hamiltonian:3-level}
\hat{H} = \hbar
\begin{pmatrix}
0 & \Omega_1/2 & 0 \\
\Omega_1/2 & \Delta & \Omega_2/2 \\
0 & \Omega_2/2 & 0
\end{pmatrix},
\end{equation}
in the basis $\{|0\rangle, |1\rangle, |2\rangle\}$, where $\Omega_1$ and $\Omega_2$ are Rabi frequencies and $\Delta$ is the common detuning. We initialize the system in state $|0\rangle$ and define a projective measurement on the target state $|2\rangle$. Thus, from equation \eqref{Eq:TFdistrib:General}, we find that the time-of-flow distribution is given by
\[
\pi(t) = \mathcal{N} \times \left|\frac{d}{dt} \, \mathrm{Tr}[\rho_t |2\rangle\langle 2|] \right|.
\]
Solving the Schrödinger equation numerically using QuTiP, we obtain an oscillatory $\pi(t)$ due to coherent transitions, {normalized over a suitable time window $[0,t_f]$. We choose the final time to be the Rabi oscillation $t_f = 2\pi/\Omega$, where $\Omega=\sqrt{\Omega_1^2+\Omega_2^2+\Delta^2}$ to maximize the population in the target state}. The standard deviation $\Delta \mathcal{T}$ of the TF distribution and the energy uncertainty $\Delta H$ of the initial state are computed and compared to the bound in equation \ref{Eq:TimeUncertaintyRelation:General}, see Figure \ref{Fig:Rabi}. 
This three-level system offers a concrete setting to examine how quantum control parameters influence the temporal profile of transitions. In particular, we observe in Fig.~\ref{Fig:Rabi} (right panel) that increasing the detuning \( \Delta \) leads to a narrower TF distribution, reflected in a decreasing uncertainty \( \Delta \mathcal{T} \). While large detuning typically reduces population transfer and slows down dynamics, it also suppresses coherent oscillations, effectively filtering out intermediate transitions and sharpening the observed timing distribution. As a result, the uncertainty product \( \Delta \mathcal{T} \cdot \Delta H \) decreases and continues to satisfy the bound~\eqref{Eq:TimeUncertaintyRelation:General}, with tightest saturation near resonance. This illustrates how the TF distribution captures the interplay between coherent driving and measurement-defined timing in discrete quantum systems.
Importantly, the product \( \Delta \mathcal{T} \cdot \Delta H \) continues to satisfy the bound~\eqref{Eq:TimeUncertaintyRelation:General} across the full range of \( \Delta \), and approaches the bound most closely in the regime of resonant or near-resonant driving $\Delta =0$. This demonstrates the sensitivity of the TF framework to interference effects and coherent control parameters, highlighting its utility in characterizing not only fundamental uncertainty but also practical limits on quantum timing precision in driven systems.

\section{Concluding Remarks.} We have introduced the concept of the \emph{Time-of-Flow} (TF) distribution as a unifying framework for addressing the problem of time in quantum mechanics. This framework encompasses both transition times in discrete and continuous spectrum observables, such as 3-level atomic systems and time-of-arrival (TOA) observables in matter-wave dynamics, providing a coherent statistical description of temporal behavior derived from projective measurements.

The TF distribution assigns a well-defined probability density to the timing of transitions into a chosen measurement subspace, constructed from a sequence of ensemble measurements that avoid Zeno inhibition. This meta-protocol is experimentally accessible and measurement-agnostic, enabling the reconstruction of the TF distribution and its statistical moments, including the standard deviation \( \Delta T \), which we interpret as the temporal uncertainty of the process.

From this distributional picture, we derived a general, context-dependent time--energy uncertainty relation, see equation \eqref{Eq:TimeUncertaintyRelation:General}. This bound holds for arbitrary projective measurements and closed dynamics, and has the advantage over conventional quantum speed limits (QSLs) in that it captures the full temporal spread of a measurable distribution. As such, it reinforces the view that time uncertainty is not a universal quantum constraint, but rather a contextual and operational property defined relative to a specific measurement protocol.

We illustrated the applicability of the TF framework in two scenarios: a driven three-level system, where coherent Rabi oscillations lead to a structured TF profile, and a free-falling quantum particle, where the TF distribution reduces to the conventional TOA distribution derived from the surface-current operator. In both cases, the predicted bound was shown to be satisfied, with experimentally relevant parameters.

The results of this work, together with the TF framework, suggest a general route for characterizing timing observables in quantum systems, where time is not a universal observable, but a context-dependent quantity whose statistical properties depend on the measurement protocol.
This operational perspective aligns naturally with experimental practice and opens the door to new precision timing diagnostics in quantum systems, such as quantum computing, quantum metrology, quantum control, and cold atom experiments.

\bibliography{References.bib}

\clearpage

\newpage

\onecolumngrid

\vspace{5mm} 

\begin{center}
\textbf{\large SUPPLEMENTARY MATERIAL}
\end{center}

\section*{Upper Bound on the TF distribution}

We derive a general upper bound on the rate of change of the expectation value of a projector $M$ under unitary evolution, expressed in terms of the energy uncertainty $\Delta H$.

Let the system evolve unitarily under the equation:
\begin{equation}
\mathcal{L}(\hat{\rho}_t) = -\frac{i}{\hbar}[\hat{H}, \hat{\rho}_t],
\end{equation}
where $\hat{\rho}_t$ is the time-dependent density operator, and $\hat{H}$ the Hamiltonian of the system. Let $\hat{M}$ be a projector, i.e., a Hermitian operator satisfying $\hat{M}^2 = \hat{M}$. We wish to bound the quantity
\begin{equation}
\left| \mathrm{Tr}(\mathcal{L}(\hat{\rho}_t) \hat{M}) \right| = \frac{1}{\hbar} \left| \mathrm{Tr}([\hat{H}, \hat{\rho}_t] \hat{M}) \right|.
\end{equation}

Using the triangle inequality and the cyclicity of the trace, we write:
\begin{align}
\left| \mathrm{Tr}([\hat{H}, \hat{\rho}_t] \hat{M}) \right|
&= \left| \mathrm{Tr}(\hat{H} \hat{\rho}_t \hat{M}) - \mathrm{Tr}(\hat{\rho}_t \hat{H} \hat{M}) \right| \\
&\leq \left| \mathrm{Tr}(\hat{H} \hat{\rho}_t \hat{M}) \right| + \left| \mathrm{Tr}(\hat{\rho}_t \hat{H} \hat{M}) \right|.
\end{align}

To each term, we now want to apply the Cauchy–Schwarz inequality. Recall that for any two Hilbert–Schmidt operators $\hat{A}$ and $\hat{B}$, we have:
\begin{equation}
|\mathrm{Tr}(\hat{A}^\dagger \hat{B})| \le \|\hat{A}\|_2 \cdot \|\hat{B}\|_2,
\end{equation}
where $\|\hat{X}\|_2^2 = \mathrm{Tr}(\hat{X}^\dagger \hat{X})$.

Let us apply this inequality with $\hat{A} = \hat{H} \hat{\rho}_t^{1/2}$ and $\hat{B} = \hat{\rho}_t^{1/2} \hat{M}$. We obtain:
\begin{equation}
\left| \mathrm{Tr}(\hat{H} \hat{\rho}_t \hat{M}) \right| = \left| \mathrm{Tr}(\hat{H} \hat{\rho}_t^{1/2} \cdot \hat{\rho}_t^{1/2} \hat{M}) \right|
\le \left( \mathrm{Tr}(\hat{\rho}_t \hat{H}^2) \right)^{1/2} \cdot \left( \mathrm{Tr}(\hat{\rho}_t \hat{M}^2) \right)^{1/2}.
\end{equation}

The same bound holds for $\left| \mathrm{Tr}(\hat{\rho}_t \hat{H} \hat{M}) \right|$ by Hermiticity of $\hat{H}$ and $\hat{M}$. Thus:
\begin{equation}
\left| \mathrm{Tr}([\hat{H}, \hat{\rho}_t] \hat{M}) \right| \le
2 \cdot \left( \mathrm{Tr}(\hat{\rho}_t \hat{H}^2) \right)^{1/2} \cdot \left( \mathrm{Tr}(\hat{\rho}_t \hat{M}^2) \right)^{1/2}.
\end{equation}

The energy variance is defined as:
\begin{equation}
\Delta H^2 := \mathrm{Tr}(\hat{\rho}_t \hat{H}^2) - \left( \mathrm{Tr}(\hat{\rho}_t \hat{H}) \right)^2.
\end{equation}

Letting $\widetilde{H} := \hat{H} - \mathrm{Tr}(\hat{\rho}_t \hat{H})$, we note that $[\widetilde{H}, \hat{\rho}_t] = [\hat{H}, \hat{\rho}_t]$ and $\mathrm{Tr}(\hat{\rho}_t \widetilde{H}^2) = \Delta H^2$. We then obtain:
\begin{equation}
\left| \mathrm{Tr}([\hat{H}, \hat{\rho}_t] \hat{M}) \right|
= \left| \mathrm{Tr}([\widetilde{H}, \hat{\rho}_t] \hat{M}) \right|
\le 2\, \Delta H \cdot \left( \mathrm{Tr}(\hat{\rho}_t \hat{M}^2) \right)^{1/2}.
\end{equation}

Finally, since $M^2 = M$ and $0 \le \mathrm{Tr}(\hat{\rho}_t \hat{M}) \le 1$ and as $\Delta H$ is time-invariant, we find:
\begin{equation}
\boxed{
\left| \mathrm{Tr}(\mathcal{L}(\hat{\rho}_t) \hat{M}) \right| \le \frac{2}{\hbar} \cdot \Delta H
}\ ,
\end{equation}
uniformly in time. 
This inequality provides a general upper bound on the transition rate into a measurement subspace defined by $\hat{M}$ in terms of the energy uncertainty $\Delta H=\sqrt{\mathrm{Tr}\left(\hat{H}^2\hat{\rho}_0\right)-\mathrm{Tr}\left(\hat{H}\hat{\rho}_0\right)^2}$.

\section*{Commutator of the Half–Space Projector $\hat{M}_{x_0}$ with the Hamiltonian}

We prove that for a one–dimensional particle of mass $m$ in a position-dependent potential $V(\hat x)$,
\begin{equation}
-\frac{i}{\hbar}\,[\hat{H},\hat{M}_{x_0}]
=
\hat{J}(x_0)\ ,
\label{eq:surface-current}
\end{equation}
where $\hat{J}(x_0) = \frac{1}{2m}\!\left(\hat p\,\delta(\hat x-x_0)+\delta(\hat x-x_0)\,\hat p\right)$, $\delta(\hat x-x_0)\equiv\ket{x_0}\bra{x_0}$, and
\begin{align}
\hat{H} =\frac{\hat p^{2}}{2m}+V(\hat x), \ \ \ \ \text{and}\ \ \ \
\hat{M}_{x_0} 
        =\int_{-\infty}^{x_0}dx'\ |x'\rangle\langle x'|\equiv\Theta(x_0-\hat x).
\end{align}
We the distribution notation as 
$$
\ket{x_0}\braket{x_0|\psi}=\psi(x_0)\ket{x_0}= \int_{-\infty}^{+\infty}dx\ \delta(x-x_0)\psi(x)\ket{x}= \int_{-\infty}^{+\infty}dx\ \delta(\hat{x}-x_0)\psi(x)\ket{x}=\delta(\hat{x}-x_0)\int_{-\infty}^{+\infty}dx\ \psi(x)\ket{x}=\delta(\hat{x}-x_0)\ket{\psi}\ ,
$$
and 
$$
M_{x_0}\ket{\psi} = \int_{-\infty}^{x_0}dx\ \psi(x)\ket{x} = \int_{-\infty}^{+\infty}dx\ \theta(x_0-x)\psi(x)\ket{x} = \int_{-\infty}^{+\infty}dx\ \theta(x_0-\hat{x})\psi(x)\ket{x}= \theta(x_0-\hat{x})\int_{-\infty}^{+\infty}dx\ \psi(x)\ket{x}=\theta(x_0-\hat{x})\ket{\psi}\ ,
$$ 
where $\ket{\psi} = \int_{-\infty}^{+\infty}\psi(x)\ket{x}$ is a test function. Thus, we have
$$
(\hat{M}_{x_0}\psi)(x)\equiv \bra{x}\hat{M}_{x_0}\ket{\psi} = \theta(x_0-x)\psi(x)\ ,
$$
where the wavefunction $\psi(x)\equiv \braket{x|\psi}$. 
Also, we have this relation
$$
\dfrac{d}{dx_0}\hat{M}_{x_0}=\ket{x_0}\bra{x_0}\ ,
$$ 
consistently with the formal relation between the two operators $\dfrac{d}{dx_0}\theta(x_0-\hat{x})=\delta(x_0-\hat{x})$. 

Because $V(\hat x)$ is a function of $\hat x$ only, we obviously have
\begin{equation}
[V(\hat x),\hat{M}_{x_0}]=0,
\end{equation}
hence, $[\hat{H},\hat{M}_{x_0}]=\frac{1}{2m}[\hat p^{2},\hat{M}_{x_0}]$. \\

Now, let us define $\hat{C}=[\hat p^{2},\hat{M}_{x_0}]$.  When the operator acts on an arbitrary wave-function
$\psi(x)=\langle x|\psi\rangle$, we obtain:
\begin{equation}
\begin{aligned}
\bigl(C\psi\bigr)(x)
&= -\hbar^{2}\partial_x^{2}\!\bigl[\Theta(x_0-x)\psi(x)\bigr]
   +\hbar^{2}\,\Theta(x_0-x)\,\partial_x^{2}\psi(x)\\[2pt]
&= \hbar^{2}\Bigl(
      2\delta(x-x_0)\,\psi'(x)
      -\delta'(x-x_0)\,\psi(x)
    \Bigr).
\end{aligned}
\label{eq:Cpsi}
\end{equation}
Using the distributional relation $\delta'(x-x_0)\psi(x)=-\delta(x-x_0)\psi'(x)$ in
\eqref{eq:Cpsi}, we find
\begin{equation}
\bigl(C\psi\bigr)(x)
= \bigl(
    \hat p\,\delta(\hat x-x_0)+\delta(\hat x-x_0)\,\hat p
  \bigr)\psi(x),
\end{equation}
which is valid for every $\psi$, hence
\begin{equation}
[\hat p^{2},\hat{M}_{x_0}] =
\hat p\,\delta(\hat x-x_0)+\delta(\hat x-x_0)\,\hat p  = \hat p\,\ket{x_0}\bra{x_0} + \ket{x_0}\bra{x_0}\,\hat p \ .
\end{equation}

By substituting the above identity,
\[
-\frac{i}{\hbar}[\hat{H},\hat{M}_{x_0}]
= -\frac{i}{\hbar}\,\frac{1}{2m}[\hat p^{2},\hat{M}_{x_0}]
= \frac{1}{2m}\bigl(
     \hat p\,\ket{x_0}\bra{x_0}+\ket{x_0}\bra{x_0}\,\hat p
  \bigr),
\]
we establishes Eq.\,\eqref{eq:surface-current}.  Taking the
expectation value in an arbitrary state $\rho(t)$:
\begin{equation}
\frac{d}{dt}\,\mathrm{Tr}\!\bigl[\hat{\rho}_t\hat{M}_{x_0}\bigr]
= -\mathrm{Tr}\bigl[\hat{\rho}_t\,\hat{J}(x_0)\bigr],
\end{equation}
demonstrating that $J(x_0):=\mathrm{Tr}\bigl[\hat{\rho}_t\,\hat{J}(x_0)\bigr]$ is the probability flux crossing the surface $x=x_0$.

\end{document}